# Reducing symmetry in topology optimization of two-dimensional porous phononic crystals

Hao-Wen Dong[1,2], Yue-Sheng Wang[1,a)], Yan-Feng Wang[1], and Chuanzeng Zhang[2]

[1]*Institute of Engineering Mechanics, Beijing Jiaotong University, Beijing 100044, China*

[2]*Department of Civil Engineering, University of Siegen, D-57068 Siegen, Germany*

In this paper we present a comprehensive study on the multi-objective optimization of two-dimensional porous phononic crystals (PnCs) in both square and triangular lattices with the reduced topology symmetry of the unit-cell. The fast non-dominated sorting-based genetic algorithm II is used to perform the optimization, and the Pareto-optimal solutions are obtained. The results demonstrate that the symmetry reduction significantly influences the optimized structures. The physical mechanism of the optimized structures is analyzed. Topology optimization combined with the symmetry reduction can discover new structures and offer new degrees of freedom to design PnC-based devices. Especially, the rotationally symmetrical structures presented here can be utilized to explore and design new chiral metamaterials.

**I. INTRODUCTION**

Understanding, modulating and controlling the wave propagation in periodic composite structures, known as phononic crystals (PnCs),[1] are important to design the novel acoustic-based devices. It is essential to determine the wave dispersion through Bragg's scattering or local resonances to achieve a range of spectra ($\omega$-space), wave vector ($k$-space), and phase properties ($\phi$-space).[2] Resulting from the destructive interference between incident elastic/acoustic waves and reflections from the scatterers, bandgaps inherently become the basis of the most applications of PnCs. Promising applications of PnCs include sound insulation,[3] sound barrier,[4] damping,[5-6] acoustic resonators,[7-8] elastic/acoustic waveguides,[9] filters,[10] frequency sensing,[11-12] acoustic mirrors,[13] switches,[14] lenses,[15] energy harvesting,[16] negative refraction,[17-18] self-collimation,[19-20] acoustic mirages,[14] rectification,[21] acoustic diodes,[22-23] and thermal conductance.[24-25]

In recent years, some researchers have been conducted and focused on topology optimization of photonic crystals (PtCs),[26-27] PnCs[28-33] and phoxonic crystals (PxCs).[34] Recent studies have shown that the topology

---

[a)] Author to whom correspondence should be addressed. Electronic mail: yswang@bjtu.edu.cn

optimization can greatly enhance the performance of PnCs, not only for bulk waves,[28-34] but also for surface waves[35] and plate waves.[36] However, because of the high computational effort and the complex optimization procedure, most topology optimizations of bandgap maximization assumed that the unit-cell has a primary high-symmetry.[26-28, 30-31, 33-34, 36] Sigmund and Hougaard[26] discovered some surprisingly simple geometric properties of optimal PtCs based on the optimization of highly symmetrical square and triangular latticed structures. Men et al.[27] designed the square and hexagonal latticed PtCs with multiple complete bandgaps by the convex conic optimization. Sigmund et al.[28] presented a systematic optimization of periodic materials and structures exhibiting phononic bandgaps. Bilal et al.[30] designed the highly square latticed solid-void PnCs with ultra-wide complete bandgaps. Dong et al. reported the symmetrical square latticed solid-solid and solid-void PnCs with large bandgaps by the single- and multi-objective optimizations in Refs.[31] and [33], respectively. They also showed how topology optimization can be used to effectively design the PxC bandgap structures and cavity.[34] Halkjer et al.[36] maximized the phononic band gaps for the infinite periodic beams and plates. Many papers about PtCs and PnCs have studied the effects of symmetry reduction of lattices[37-40] or unit-cells[41] on the bandgaps. Wang et al.[37] investigated the effects of shapes and orientations of scatterers and lattice symmetries on PtC bandgaps. Anderson et al.[38] also discussed the lattice symmetry reduction on PtC bandgap structures. Malkova et al.[39,40] presented the symmetrical perturbation analysis of PtCs and predicted the band spectrum evolution. Kuang et al.[41] studied the effects of the shapes and symmetries of scatterers on PnC bandgaps. They all found that the symmetry reduction can increase the width of the bandgaps or create new gaps. Therefore, it is necessary to identify the structural form of a PnC which can possess the largest bandgaps for a given material combination. Some researchers have obtained interesting optimized topologies based on the asymmetric assumption.[29, 32, 42] Preble et al.[42] found that a PtC with low symmetry can open a large TE bandgap. Gazonas et al.[29] performed optimization with asymmetric assumption and obtained the symmetrical PnCs in the coarse grids. Dong et al.[32] designed the asymmetric solid-solid PnCs with ultra-large bandgap widths by topology optimization. However, in view of the limitations of single-objective optimization, there are still many open and intriguing questions, e.g. what is the effect of the symmetry reduction on the optimized PnCs? What is the nominal optimal PnC for the given material phases and lattice symmetry?

In this paper, based on the multi-objective topology optimization, we present a comprehensive study about the effect of the symmetry reduction of the unit-cell on the optimized porous PnCs. The materials considered in this study are the same as in our previous article[33] to show the improvement of the solutions. The finite element method

(FEM) is used to calculate the band structures. The multi-objective optimization procedure is performed by the non-dominated sorting genetic algorithm II.[43] We perform the topology optimizations for PnCs with the rotationally symmetrical or asymmetrical unit-cells in the square lattice and the triangle-symmetrical or rotationally symmetrical unit-cells in the triangular lattice, and compare them with the square-symmetrical case in our previous paper.[33]

In Section II, we introduce the multi-objective optimization of PnCs for simultaneously bandgap maximization and mass minimization. We present and discuss the optimized results for both square and triangular lattices in Sections III(A) and III(B), respectively. In Section IV, we analyze the physical mechanisms of the optimized structures with different symmetry assumptions. Finally, we present a summary in Section V.

## II. MULTI-OBJECTIVE OPTIMIZATION METHOD FOR PHONONIC CRYSTAL DESIGN

The optimization problem herein is based on the PnC with holes in silicon.[30] The harmonic wave equation in an elastic solid is given by

$$(\lambda + 2\mu)\nabla(\nabla \cdot \mathbf{u}) - \mu\nabla \times \nabla \mathbf{u} + \rho\omega^2 \mathbf{u} = 0, \tag{1}$$

where $\lambda$ and $\mu$ are the Lame constants; $\rho$ is the mass density; $\mathbf{u}$ is the displacement vector; $\omega$ is the angular frequency; and $\nabla$ is the gradient operator. Here in the present paper, we will consider two-dimensional case in which independent in-plane and out-of-plane (or anti-plane) wave modes propagate in solids. It is well known that the bandgap properties (e.g. the frequency regions) for in-plane and anti-plane wave modes are different. Furthermore, if we consider the propagation of both in-plane and out-of-plane waves simultaneously, we will have bandgaps which are completely different from those for either in-plane mode or out-of-plane mode. So in this paper we will present optimization of PnCs for the three situations: 1) only in-plane wave mode propagating in PnCs, 2) only out-of-plane wave mode propagating in PnCs, and 3) full wave mode, i.e. both in-plane and out-of-plane wave modes propagating simultaneously in PnCs.

To obtain the band structures for a PnC unit-cell we consider the Bloch conditions to the solution of Eq. (1) in the form $\mathbf{u}(\mathbf{r}) = e^{i(\mathbf{k}\cdot\mathbf{r})}\mathbf{u}_\mathbf{k}(\mathbf{r})$, where $\mathbf{u}_\mathbf{k}(\mathbf{r})$ is a periodic function of the spatial position vector $\mathbf{r}$ with the same periodicity as the structure, and $\mathbf{k}=(k_x, k_y)$ is the Bloch wave vector. We use FEM to calculate the whole dispersion relations ($\mathbf{k}$-$\omega$) by the ABAQUS/Standard eigen-frequency solver Lanzcos[31-34] combined with the Python scripts. Because ABAQUS cannot directly solve the eigenvalue equations in complex form[44], we write the discrete form of the eigenvalue equations in the unit-cell in the form of[44]

$$\left( \begin{bmatrix} \mathbf{K}_R & -\mathbf{K}_I \\ \mathbf{K}_I & \mathbf{K}_R \end{bmatrix} - \omega^2 \begin{bmatrix} \mathbf{M}_R & -\mathbf{M}_I \\ \mathbf{M}_I & \mathbf{M}_R \end{bmatrix} \right) \begin{bmatrix} \mathbf{u}_R \\ \mathbf{u}_I \end{bmatrix} = 0, \tag{2}$$

where the subscripts $R$ and $I$ denote the real and imaginary parts of the unit-cell model in ABAQUS respectively.

The Bloch conditions at the boundary nodes of the unit-cell are implemented in ABAQUS as[31]

$$\begin{bmatrix} \mathbf{u}_R(\mathbf{r}) \\ \mathbf{u}_I(\mathbf{r}) \end{bmatrix} = \begin{bmatrix} \cos(\mathbf{k} \cdot \mathbf{a}) & \sin(\mathbf{k} \cdot \mathbf{a}) \\ -\sin(\mathbf{k} \cdot \mathbf{a}) & \cos(\mathbf{k} \cdot \mathbf{a}) \end{bmatrix} \begin{bmatrix} \mathbf{u}_R(\mathbf{r}+\mathbf{a}) \\ \mathbf{u}_I(\mathbf{r}+\mathbf{a}) \end{bmatrix}. \tag{3}$$

Solving Eqs. (2) and (3) with ABAQUS, then the band structures can be obtained by scanning the edges of the first Brillouin zone.

The goal of this paper is to explore the effect of the symmetry reduction of the unit-cell on the optimized PnCs. We perform both the multi-objective and the single-objective optimization to show the variation of the optimized structures. To arrive at the low-cost and high performance PnCs, the multi-objective optimization problem (MOOP) is formulated and solved for searching the structures with the maximal relative bandgap width (BGW) and the minimal mass[33]. Thus, the multi-objective optimization problem (MOOP) can be formulated as

Find: $x_i$ ($i=1, 2, …, M$),

$$\text{Maximize: } f_n(\Sigma) = \frac{\Delta\omega^2(\Sigma)}{\omega^2_c(\Sigma)} = 2 \frac{\min_{\mathbf{k}} : \omega^2_{n+1}(\Sigma,\mathbf{k}) - \max_{\mathbf{k}} : \omega^2_n(\Sigma,\mathbf{k})}{\min_{\mathbf{k}} : \omega^2_{n+1}(\Sigma,\mathbf{k}) + \max_{\mathbf{k}} : \omega^2_n(\Sigma,\mathbf{k})}, \tag{4}$$

$$\text{Minimize: } m = \sum_{i=1}^{M} \rho_i V_s, \tag{5}$$

Subject to: Eqs. (2) and (3) (5a)

$\quad\quad\quad\quad f_n > f_0,$ (5b)

$\quad\quad\quad\quad \min_{\Sigma}(b) > b^*,$ (5c)

$\quad\quad\quad\quad x_i = 0 \text{ or } 1,$ (5d)

where $\rho_i$ is the mass density of the element in the design domain and declares the absence (0) or presence (1) of a solid element; $M$ indicates the number of elements used to discretize the design domain; $f_n$ is the "relative BGW" between the $n$th and ($n$+1)th bands ($n$=1, 2, …, 10); $\Sigma$ denotes the topological material distribution within the unit-cell; $m$ is the "mass" of the unit-cell; $V_s$ is the volume ratio of the solid; $f_0$ is the prescribed value which is set as 0.8 through numerical tests to yield more solutions with large BGW;[33] $b$ is the geometrical width of the connection which must be larger than a prescribed value $b^*$ (we take $b^*=a/30$ as in Ref. [33]). The geometrical constraint (5c) is used to control the size of the minimal connection to overcome the mesh-dependence for porous PnCs.[33] We will consider four types of unit-cells during the following topology optimizations, i.e. the unit-cells with rotational symmetry and asymmetry in square lattices, and triangular symmetry and rotational symmetry in triangular lattice. A low symmetry of a unit-cell implies that more **k**-vectors are needed in band structure calculations, and thus more computing time. Therefore highly effective techniques to get the band structures are necessary.

The optimization problem of Eqs. (4)-(5d) can be efficiently solved by using the fast non-dominated sorting-based genetic algorithm II (NSGA-II),[43] which is one of the most popular and efficient algorithm for multi-objective optimization problems. For comparison, we also present the results from the single objective optimization problem (SOOP) for maximal relative BGW between the two prescribed adjacent energy bands. Compared with the MOOP, the SOOP has only one objective function, i.e. the relative BGW, and therefore it is easier and needs less computational efforts. We solve the SOOP by using the genetic algorithm (GA),[31-34] which has already been used in the design of structures with a very large search space or high dimension. In general, GA requests a large number of generations to convergence, so it is perhaps the best method for the problems having a low computational cost for the fitness evaluation. Both algorithms of GA and NSGA-II are based on the principle of the biological evolution, and therefore their iterative procedures are similar. Figure 1 shows these two procedures and can be briefly described as follows:

1) Start with a mesh $N \times N$ and a random initial parent population $P_0$ of $N_p$ individuals. Each individual represents a unit-cell structure.

2) To evaluate the fitness of every individual, all structures are computed for the optimization objectives. For GA the relative BGWs are computed, and for NSGA-II both the relative BGWs and mass are calculated.

3) For NSGA-II, the non-dominated sorting operation is needed to get the non-dominated relationship among various solutions.

4) Perform the genetic operations, including the selection, crossover, mutation and local search. The local search is used to optimize the structure to some extent, see Ref. [33].

5) For GA, after the genetic operations, the algorithm yields the offspring population $Q_n$ which will be taken as the parent $P_n$ of the new generation. However, for NSGA-II, the non-dominated sorting and crowding distance sorting operations are adopted to get the final offspring $Q^*_n$ based on $P_n$ and $Q_n$.

6) The iteration will end when the prescribed maximal generation is reached. Otherwise, the next generation will be considered (i.e., $n=n+1$) and the procedure returns to step (2).

More detailed descriptions and discussions about these two algorithms can be found in our previous works[31-34] and Ref. [43].

**FIG. 1.**

## III. NUMERICAL RESULTS AND DISCUSSIONS

The above described optimization method can be applied to an arbitrarily shaped lattice. Next we will present the computational results for the 2D square-latticed and triangle-latticed PnCs made of silicon[30, 33] with vacuum holes. Although silicon is generally anisotropic, here we select the isotropic silicon with $\rho$=2330 kgm$^{-3}$, $\lambda$=85.502 GPa, $\mu$=72.835 GPa and $c_{t,\text{silicon}}$=5591 ms$^{-1}$ as in Refs. [30, 33] for the reason of comparison between the present results and the previous ones.[33] In Sec. III(A), we will discuss the influence of the anisotropy on the optimized structure. The algorithm parameters of the binary-coded NSGA-II are the population size $N_p$=30, the crossover probability $P_c$=0.9, and the mutate probability $P_m$=0.02. We initialize our procedure based on a coarse grid 30×30 and obtain the optimized solutions ($\Sigma_1, \Sigma_2, …, \Sigma_{30}$) after 1000 evolutionary generations. Then, these Pareto-optimal solutions are mapped to the fine grid 60×60 in resolution and used as the initial population in the new-run of the optimization procedure. The numbers of the design variables of a design domain with the rotational symmetry and asymmetry are $2^{900}$ and $2^{3600}$, respectively. The numerical tests show that the optimization procedure converges and the final optimization results ($\Sigma_1^*, \Sigma_2^*, …, \Sigma_{30}^*$) are obtained after 6000 generations. We typically need 210000 fitness evaluations (i.e. the band structure calculations for any wave mode) during the whole optimization process. All computations were performed on a Linux cluster with Intel Xeon E5-2660 @2.20 GHz. And each run in our procedure was obtained in about 168 hours with 30 CPUs. For the GA adopted in this analysis, we use the following algorithm parameters: the population size $N_p$=20, the crossover probability $P_c$=0.9, the mutate probability $P_m$=0.02, and the maximal generation number $M_g$=2000. The same "coarse to fine"[33] optimization strategy is used as in the NSGA-II.

For all structures shown below, we consider various prescribed symmetries, i.e. the square-symmetry, rotational symmetry and asymmetry in the square lattices, and triangle-symmetry and rotational symmetry in the triangular lattices, respectively. Note that all the corresponding results for the square-symmetry can be found in our previous work.[33] We also consider different types of wave modes for the bandgap optimization, i.e. the out-of-plane mode, in-plane mode, and full wave mode of both out-of-plane and in-plane waves. The study about the effect of the symmetry reduction of the unit-cell on optimizing bandgaps enables us to choose the most appropriate design. Below we present some representative results to illustrate the optimized structures with different unit-cell's symmetries in both square and triangular lattices for different wave modes. The physical mechanisms and the potential applications are also discussed.

**A. SQUARE LATTICE**

The optimization procedure should start from the random design population due to the complicated nature of the solution space. Although the specially designed "seed" structure is helpful for accelerating the optimization procedure for the SOOP,[31] the artificial design will lead to local optima and even make the optimization procedure difficult to evolve for the MOOP. Moreover, because of the large number of the design variables ($2^{900}$ for the rotational symmetry or $2^{3600}$ for the asymmetry), the MOOP needs more evolutionary generations to get the trade-off relationship between the relative BGW and mass. It is well known that the optimization results are sensitive to the genetic parameters, such as the population size, crossover probability and mutate probability. So, we repeatedly computed the optimization problems and obtained the relatively stable and convergent results.

The non-convexity of the optimization problem makes it difficult to prove that our optimized solutions are optimal. However, we can take our present solutions as the effectively optimized solutions. As shown in Fig. 2, we illustrate the effective optimization results from the MOOP and SOOP for the out-of-plane wave mode. The scattered hollow diamonds, circles and squares are the Pareto-optimal solutions for the square-symmetry, rotational symmetry and asymmetry, respectively. The optimal solutions of the SOOP for these three cases are presented by the scattered solid symbols in Fig. 2. It can be seen from this figure that for both SOOP and MOOP, the optimized solutions of the asymmetry are the best, and those of the rotational symmetry are the second best. From the perspective of the optimization problem, the search space will increase significantly with the reduction of the unit-cell's symmetry. So, the design domain with the asymmetrical unti-cell has the most abundant solution space and should be nearest to the nominal optimal solutions in the bandgap engineering for the 2D porous PnCs. In other words, the asymmetrical unit-cell is maybe the most excellent design for improving the bandgap width. Analogously, the unit-cell with rotational symmetry is better than the square-symmetrical one. The Pareto-curve identifies a subset of structures with satisfactory values of both bandgap width and mass. If a large $f_1$ (or big $m$) is of the predominant importance, then the designs $B_r$ and $B_a$ are two good choices; if a small mass (or small $f_1$) is required, the designs $A_r$ and $A_a$ become attractive candidates. Three solutions of the SOOP have similar relative BGWs as those of the

MOOP and locate on the Pareto-curves. This means that they are near-optimal structures for the bandgap width maximization. In view of the complexity of the optimization problem, unfortunately we cannot theoretically prove our optimized solutions are optimal, so the optimized solutions in this paper are called near-optimal ones. Hence, this illustrates the effectiveness of both multi-objective and single-objective optimization methods in this paper.

**FIG. 2.**

Figure 2 also presents the representative designs of the optimized solutions with the rotational symmetry ($A_r$ and $B_r$) and asymmetry ($A_a$ and $B_a$) to show their geometrical features. In fact, all optimized structures on the same curves have a similar topology. Every solution on the Pareto-optimal curve is the near-optimal one for any objective. Despite of the complexity of the optimized structures, the decision maker can change the structure to some extent to get the simple and effective one. This should be the important meaning of the topology optimization for the PnCs design. According to two optimized structures $A_r$ and $B_r$, it is interesting to observe that they have the rotationally symmetrical solid lumps with the rotationally symmetrical connections. Our previous study[33] has shown that the porous PnCs with large solid lumps and narrow connections can open large bandgaps. However, for the out-of-plane wave mode, the relative BGW of rotated design $B_r$ is 15% larger that of the near-optimal structure $S_s$ in Ref. [33]. According to $A_r$, $B_r$, $A_a$ and $B_a$, the bandgap width increases as the solid lumps become larger.

Because no symmetry constraint about the unit-cell model is prescribed, the optimization procedure for the optimized asymmetrical structures leads to the "ugly" and complicated topologies, see Fig. 2. But, we can observe that the optimized structures are also composed of the concentrated solid lumps and narrow connections, just like the structures with square-symmetry and rotational symmetry in Fig. 2. Compared with $B_r$, the design $B_a$ has a larger relative BGW and a smaller mass. Moreover, besides the smaller mass, the relative BGW of the design $B_a$ is 1.354, which is 30.95% larger than that of the near-optimal structure $S_s$ in Ref. [33]. Therefore, for the out-of-plane wave mode, we can conclude that the final optimal structure for 2D porous PnCs generally favors the asymmetrical topology. Of course, with the reduction of the unit-cell's symmetry, the optimized structures will have larger bandgap widths and smaller masses. We expect that the same tendency can be found for the other symmetries.

It is important to note that, no matter what symmetrical unit-cell is considered, the topology optimization in this paper is performed based on the assumption that the discretized finite elements on the opposite edges of the unit-cell model in ABAQUS are the same. We adopt this approach in order to easily apply the Floquet-Bloch wave conditions in the discrete optimization model.

We also perform the topology optimization for the in-plane wave mode. Figure 3 shows the solutions for optimizing the third bandgap which locates between the third and forth bands and is most easily generated for the in-plane wave mode. For the three symmetries, both the multi-objective and single-objective optimization results are presented. Overall the solutions are smaller than those for the out-of-plane mode as shown in Fig. 2. The rotationally symmetrical and asymmetrical solutions are better than the square-symmetrical ones. Comparing Figs. 2 and 3, we find that the symmetry reduction of the unit-cell in a square lattice has a more significant influence on the optimized bandgaps for the out-of-plane mode than for the in-plane mode.

**FIG. 3.**

The representative designs for the rotationally symmetrical and asymmetrical cases for the in-plane wave mode are displayed in Fig. 3. Similar to the previous results for the out-of-plane mode, we note that all the optimized rotationally symmetrical structures have the rotational lumps and narrow curved connections. The optimized asymmetrical structures have the irregular lumps and connections. For both rotational and asymmetrical designs, the bigger the centered solid lumps are, the larger the relative BGW is. It is important to note from Fig. 3 that the two nearby lumps of the design $B_r$ nearly contact. The same geometrical feature can be observed for the design $B_a$ in Fig. 3. Obviously, this topology should be the limiting state for optimizing the bandgaps of the in-plane wave mode. So it is difficult to find out a structure which has the larger bandgap size and mass than those of the design $B_a$ with a relative BGW of 1.636 and mass of 0.789. This result in the "faulty" multi-objective solutions, i.e., the Pareto-curve in Fig. 3 for the asymmetry, does not "contain" that for the rotational symmetry.

Besides, the topology optimization shows its strong space searching capacity again. The optimized design $S_r$ has a 15.05% larger bandgap width than that of the near-optimal design $S_s$ in Ref. [33], and the asymmetrical design $S_a$ has a much larger relative BGW. In view of the no constraints assumption on the uniti-cell, the optimized asymmetrical structure $S_a$ should be the most excellent design for the bandgap engineering of the 2D porous PnCs. In addition, the rotational design $B_r$ is also a good choice owing to its sufficiently large relative BGW and rotating feature, especially when the mass is more concerned.

To consider all wave types propagating in PnCs, we also present the multi-objective optimization solutions for the full wave modes in Fig. 4. Because the complete bandgaps are determined by the bandgaps for both the *xy*- and *z*-polarizations, the optimized multi-objective results have smaller increases with the symmetry reduction compared with the results for the pure out-of-plane or pure in-plane modes. However, we still observe that the optimized designs with a lower symmetry can possess larger bandgaps and smaller mass. From Fig. 4, it is important to note that the rotationally symmetrical and asymmetrical results obviously offer more solutions than the square-symmetrical ones. This again proves that the symmetry reduction of the unit-cell will make a much richer solution set.

**FIG. 4.**

Figure 4 shows the representative designs as well. The same topology feature of the designs can be found for the rotationally symmetrical and asymmetrical cases. Comparing the rotational results for the three types of the wave modes, i.e. Figs. 2, 3 and 4, we find that the only difference is the geometry of the center lump. So, we can get the structure with the desired property by adjusting the geometry and size of the solid lump. To the best knowledge of the authors, the optimized asymmetrical structure $S_a$ in Fig. 5(b) is the best solution ever reported for the complete bandgap whose relative BGW is 22.85% larger than that of the structure $S_s$ in Ref. [33]. The optimized rotational structure $S_r$ in Fig. 5(a) also has an excellent nature in the bandgap size. The first complete bandgap lies between the third and fourth bands for the *xy*-polarization, and between the first and second bands for the *z*-polarization. The corresponding relative BGW of 1.223 is quite remarkable, and is 14.51% larger than that of the structure $S_s$ in Ref. [33]. In general, compared with the asymmetrical design, we prefer the optimized rotationally symmetrical structure due to its relatively simple topology. Moreover, we find that the designs $S_r$ and $S_a$ can open multiple bandgaps which are of considerable interest. Because this property can be utilized to design the PnC-based devices operating in a wider range of forbidden frequencies, such as the multi-band-pass filters.

**FIG. 5.**

In order to find out the effect of the material anisotropy on the optimized solutions, we present the results for the anisotropic silicon whose material parameters are selected as: $\rho$=2331 kgm$^{-3}$, $C_{11}$=16.57×10$^{10}$ Pa, $C_{12}$=6.39×10$^{10}$Pa, and $C_{44}$=7.962×10$^{10}$ Pa.[34,56] Figure 6 shows the optimized solutions with square-symmetry and rotational symmetry for the third bandgap of the in-plane wave mode. The material orientation is defined as $\theta$, with $\theta=0°$ corresponds to the case where the material principle axes are parallel/perpendicular to the symmetrical directions of the unit-cell. It is note that the material symmetry reduction has significant effects on the bandgap maximization and mass minimization. But all optimized topologies of solutions are similar to those with isotropic material parameters. For both cases of $\theta=0°$ and $\theta=30°$ the rotational symmetrical structures are superior to the square-symmetrical ones in view of bandgaps. Meanwhile, the gaps between the rotational symmetry and square-symmetry are larger for the lower material symmetry cases ($\theta=30°$). Besides, for the square-symmetry assumption, the solutions of $\theta=30°$ are dominated by those of $\theta=0°$. However, for the rotational symmetry, the solutions of $\theta=30°$ are nearly coincident with those of $\theta=0°$ in the area of large bandgaps. In the area of smaller bandgaps, the optimized solutions with $\theta=30°$ are better. Therefore, reducing the material symmetry is also useful for bandgap engineering. With the lower material symmetry, the structural symmetry reduction can easily get the larger improvement.

**FIG. 6.**

## B. TRIANGLE LATTICE

In the following subsection, we investigate the topology optimization of PnCs with the triangle-symmetry and rotational symmetry in the triangular lattices. As far as the authors know, no topology optimization of porous PnCs in the triangular lattice has ever been reported in literature. So, the following optimized solutions will firstly provide a guide for the bandgap maximization in the triangular lattices. Now we first consider the out-of-plane wave mode. Analogously, we present the Pareto-optimal solutions for two symmetry assumptions and the corresponding representative designs in Fig. 7. We observe that both designs $A_t$ and $A_r$ have very small masses, making them good choices for ultra-light acoustic functional materials. It is particularly interesting to remark that the results for the triangle-symmetry are a little bit better than those for the rotational symmetry. This is quite different from the case for the square lattice in Fig. 2, which shows that the optimized results with a low symmetry have a better performance than those with a high symmetry. However, we would like to emphasize that the optimization for the rotationally symmetrical structures excludes the triangle-symmetry. If we include the triangle-symmetry in the optimization of the rotationally symmetrical structures, we can find that the optimized structure with the triangle-symmetry is the best. To support this argument, we will further analyze and discuss the physical mechanisms by comparing the near-optimal structures $T_t$ and $T_r$ in the following Sec. IV.

**FIG. 7.**

We now turn to the bandgap maximization for the in-plane wave mode, see Fig. 8. Unlike the results for the out-of-plane wave mode in Fig. 7, the results in Fig. 8 show that the whole Pareto-curve of the optimized rotational designs is superior to the triangle-symmetry case, and the gap between the two cases increases as the objective function $f_3$ gets larger. Obviously, for the in-plane wave mode, the symmetry reduction of the unit-cell can improve the bandgap width with the same mass. Moreover, similar to the results for square lattices, the optimized rotational structure can open a bandgap with a small mass. However, the rotational symmetry is inimical to the generation of a bandgap when the structure has a large mass. According to Fig. 8, both the near-optimal solutions for the triangle-symmetry and the rotational symmetry just occur on the Pareto-curves. Comparing the two near-optimal solutions, the optimized rotational one has the best performances regarding the bandgap width for the same mass.

**FIG. 8.**

The present optimization method is able to optimize the complete bandgaps. As shown in Fig. 9, the optimized rotational Pareto-optimal curve can only find a little better solutions in the range with a small mass, and can find more optimized solutions in the range with a larger mass. This means that the optimized designs with the triangle-symmetry are good enough. This peculiarity is attributed to the fact that, for the out-of-plane wave, the optimized solutions for the triangle-symmetry are optimal in the triangular lattices, see Fig. 7. On the other hand, the symmetry reduction can help us to find out more effective solutions. It is worth noting that the rotational case for full wave modes in Fig. 9 has very similar designs to those for the in-plane wave mode in Fig. 8. This suggests that the optimal structures with the triangle-symmetry mainly depend on the opening of bandgaps for the in-plane wave mode. We also observe, however, that the rotationally symmetrical structure can obtain an ultra-wide bandgap width with a large mass. Whereas, with the same large mass, the triangle-symmetrical one cannot open bandgaps anymore.

The near-optimal structure $T_t$ has a bandgap width of 1.08, which is larger than that of the structure $T_r$. So, for the single objective optimization, the optimized structure with the triangle-symmetry is the best. We emphasize here again that no other porous PnCs with a so large bandgap size in triangular lattices have been reported yet in literature. As shown in Fig. 10, the structure $B_t$ can open multiple and complete gaps. Like the work of Ref. [27], due to the prohibition of waves at multiple frequencies, this PnC may be useful for the suppression of resonances at certain frequencies.

**FIG. 9.**

**FIG. 10.**

## IV. PHYSICAL MECHANISM ANALYSIS

As shown in Sec. III, the optimized structures in the square and triangular lattices have the excellent performances in the bandgap size. Moreover, for all wave modes, all optimized structures have the big solid lumps and narrow connections. To show their physical mechanism, we calculate the band structures and the corresponding vibration modes at the bandgap edges for the optimized representative structures in Figs. 2, 3, 7 and 8, and they are presented in Fig. 11. For the out-of-plane wave, the results for the optimized structures $B_r$ in Fig. 2, as well as $T_t$ and $T_r$ in Fig. 7 are shown in Figs. 11(a)-(c) respectively. Figures 11(d) and 10(e) present the results for the optimized structures $S_r$ in Fig. 3 and $B_t$ in Fig. 8 for the in-plane waves.

**FIG. 11.**

The optimized structures can be modeled as mass-spring systems, i.e., the solid lumps act as masses and the bars as the springs.[59] It is seen from the modes $O_1$, $O_3$ and $O_5$ that the lumps vibrate strongly and the bars vibrate slightly for the lower-edge modes. However, for the upper-edge modes $O_2$, $O_4$ and $O_6$, the situation is the reverse. According to mode $I_1$ in Fig. 11(d), the four lumps rotate with the bars vibrating. Conversely, the bars of mode $I_2$ rotate but the lumps hardly move. The similar vibration properties can be observed in modes $I_3$ and $I_4$, i.e., for the lower-edge the lumps have a translation motion and the connections have minute vibrations in different directions for different structures, while for the upper-edge the vibrations are just reversed. So, the generation of the optimized bandgap for the optimized structures is caused by the local vibrations. This means that the larger effective mass of the lump and/or smaller effective stiffness of the connection will lead to a larger upper eigen-frequency and a smaller lower eigen-frequency. Fortunately, the symmetry reduction of the unit-cell can help the topology optimization to find out as much as possible such structures with a sufficiently large mass and a sufficiently narrow connection.

However, some quite unusual results are observed in Figs. 11(b) and 11(c). In Fig. 7 of Sec. III, our optimization results have indicated that, for both multi-objective and single-objective optimizations, the optimal solutions with the triangle-symmetry are better than those with the rotational symmetry for the out-of-plane mode. In fact, the optimized structure in Fig. 11(c) has four particular connections, even though the structures in Figs. 11(b) and 11(c) have the similar topology, band structures and vibration modes. Obviously, these connections are wider than those in Fig. 11(b). Then, the structure in Fig. 11(b) has a larger upper eigen-frequency and a smaller lower eigen-frequency. Because a structure in triangular lattices with the triangle-symmetry can provide the narrowest connections than that with any other possible symmetry, the optimized structures with the triangle-symmetry should be the optimal solutions for the out-of-plane wave.

**V. CONCLUSION REMARKS**

In pursuit of the nominal optimal porous PnCs, in this paper, we have demonstrated the effect of the symmetry reduction of the unit-cell on the optimized solutions for the MOOP, i.e., the simultaneously maximal bandgap size and minimal mass. For comparison, the optimized solutions for the SOOP are also presented. The unit-cell's symmetries include the rotational symmetry and asymmetry in the square lattice, and the triangle-symmetry and rotational symmetry in the triangular lattice. From the results presented in this paper, conclusions can be drawn as follows:

1) We have numerically demonstrated that, for the out-of-plane, in-plane and full wave modes, the symmetry reduction of the unit-cell will result in the better Pareto-optimal solutions with the bigger relative bandgap widths and smaller masses. Consideration of the symmetry reduction will provide a guiding principle towards the design of PnCs. We should point out that our optimization procedure can be also extended to unit-cells with other types of symmetries or lattice forms.

2) From our investigations, we have shown that the combination of the unit-cell's symmetry reduction and the topology optimization can be an effective way to obtain the ultra-wide bandgaps with ultra-small mass simultaneously. Moreover, the principle of the optimized rotational PnCs can be extended for exploring and designing the chiral metamaterials. Though all optimized structures in this paper show the common feature in geometry (large lumps and narrow connections), a big difference can also be found for the different symmetry assumptions and lattice types. Therefore, the numerical design by the topology optimization is still crucial to the bandgap engineering of PnCs.

3) For the square lattice and in any wave mode, the optimal structures with an asymmetrical unit-cell should be the best choices. For the triangular lattice, the optimized rotational structures show more improvements compared with the triangle-symmetrical case in the in-plane wave mode. However, for both out-of-plane and full wave modes, the rotational structures are hardly fully inferior compared to the triangle-symmetrical ones. Specially, the optimized structures with the triangle-symmetry should be the optimal solutions for the out-of-plane wave.

4) In addition to structural symmetry, the material symmetry reduction of the anisotropic material also affects the multi-objective solutions significantly. Nonetheless, the optimized topologies of solutions are similar to those with isotropic material parameters. With the lower material symmetry, the structural symmetry reduction can easily get the larger improvement of bandgaps.

Because of the non-convexity of the multi-objective optimization problems and limited computing capacity, we should point out here that the multi-objective optimization results presented in this paper are near-optimal rather than perfect. In the future work, we will extend the adopted optimization method to get the better and perfect Pareto-optimal solution set. More importantly, we should also consider the manufacturing and fabrication related issues and find out more realistic and valuable optimized structures for practical designs. So, we believe that the consideration of the symmetry reduction and the fabrication robustness in the topology optimization can help us to reach the target of the PnC bandgap maximization and even manipulating special wave propagation[45-47]. In addition, we can also use the topology optimization to design PnC bandgaps with the ultra-low mid-frequencies for low frequency sound insulation.[60] Moreover, the topology optimization combined with the symmetry reduction may be suitable for the PxC bandgap engineering[34, 48-58] to construct more PxCs with a strong photon-phonon interaction. This interesting problem is the subject of our ongoing research.

## ACKNOWLEDGMENTS

This work is supported by the Fundamental Research Funds for the Central Universities (2015YJS125). The second author is also grateful to the support by the National Natural Science Foundation of China under Grant No. 11532001. The paper was completed during the stay of the first author in Germany under the support from the Chinese Scholarship Council (CSC) and the German Academic Exchange Service (DDAD) through the Sino-German Joint Research Program (PPP).

**FIGURE CAPTIONS**

**FIG. 1.** Iterative procedures of GA and NSGA-II.

**FIG. 2.** Multi-objective solutions for the square-symmetry (scattered hollow diamonds), rotational symmetry (scattered hollow circles) and asymmetry (scattered hollow squares) in the square lattices with simultaneously maximal relative BGWs and minimal mass for the first bandgap $f_1$ of the out-of-plane wave mode. The results for the square-symmetry are from Ref. [33]. The near-optimal solutions of the SOOP with the square-symmetry (solid diamond), rotational symmetry (solid circle) and asymmetry (solid square) are also shown. The representative designs $A_r$, $B_r$, $A_a$ and $B_a$ are presented as well.

**FIG. 3.** Multi-objective solutions for the square-symmetry (scattered hollow diamonds), rotational symmetry (scattered hollow circles) and asymmetry (scattered hollow squares) in the square lattices with simultaneously maximal relative BGWs and minimal mass for the third bandgap $f_3$ of the in-plane wave mode. The results for the square-symmetry are from Ref. [33]. The near-optimal solutions of the SOOP with the square-symmetry (solid diamond), rotational symmetry (solid circle) and asymmetry (solid square) are also shown. The representative designs $A_r$, $B_r$, $A_a$ and $B_a$ are presented as well.

**FIG. 4.** Multi-objective solutions for the square-symmetry (scattered hollow diamonds), rotational symmetry (hollow circles), asymmetry (hollow squares) in square lattices with simultaneously maximal relative BGWs and minimal mass for the first complete bandgap $f_1$ of the full wave modes. The results for the square-symmetry are from Ref. [33]. The near-optimal solutions of the SOOP with the square-symmetry (solid diamond), rotational symmetry (solid circle) and asymmetry (solid square) are also shown. The representative designs $A_r$, $B_r$, $A_a$ and $B_a$ are presented as well.

**FIG. 5.** Optimized 4×4 crystal structures $S_r$ (a), $S_a$ (b) in Fig. 4, their band structures and the corresponding first irreducible Brillouin zone. The solid and dashed lines represent the bands for the in-plane and out-of-plane modes, respectively. The normalized frequency $\Omega=\omega a/2\pi c_t$ (with the transverse wave velocity of silicon and the lattice constant $a$) is used.

**FIG. 6.** Multi-objective solutions for the square-symmetry (circles and pentagons) and rotational symmetry (squares and triangles) in the square lattices with anisotropic material parameters for the third bandgap $f_3$ of the in-plane wave mode. Different material orientations ($\theta=0°$ and $\theta=30°$) are considered.

**FIG. 7.** Multi-objective solutions for the triangle-symmetry (solid lower triangles) and rotational symmetry (solid upper triangles) in the triangular lattices with simultaneously maximal relative BGWs and minimal mass for the first bandgap $f_1$ of the out-of-plane wave mode. The near-optimal solutions of the SOOP with the triangle-symmetry (hollow lower triangle with cross) and rotational symmetry (hollow upper triangle with cross) are also shown. The representative designs $A_t$, $B_t$, $A_r$ and $B_r$ are presented as well.

**FIG. 8.** Multi-objective solutions for the triangle-symmetry (solid lower triangles) and rotational symmetry (solid upper triangles) in the triangular lattices with simultaneously maximal relative BGWs and minimal mass for the third bandgap $f_3$ of the in-plane wave mode. The near-optimal solutions for the SOOP with the triangle-symmetry (hollow lower triangle with cross) and rotational symmetry (hollow upper triangle with cross) are also shown. The representative designs $A_t$, $B_t$, $A_r$ and $B_r$ are presented as well.

**FIG. 9.** Multi-objective solutions for the triangle-symmetry (solid lower triangles) and rotational symmetry (solid upper triangles) in the triangular lattices with simultaneously maximal relative BGWs and minimal mass for the first complete bandgap $f_1$ of the full wave modes. The near-optimal solutions of the SOOP with the triangle-symmetry (hollow lower triangle with cross) and rotational symmetry (hollow upper triangle with cross) are also shown. The representative designs $A_t$, $B_t$, $A_r$ and $B_r$ are presented as well.

**FIG. 10.** Optimized 4×4 crystal structure $B_t$ in Fig. 9, its band structures and the corresponding first irreducible Brillouin zone.

**FIG. 11.** Optimized representative structures and their band structures for the out-of-plane wave: (a) $B_r$ in Fig. 2, (b) $T_t$ in Fig. 7, and (c) $T_r$ in Fig. 7; and in-plane waves: (d) $S_r$ in Fig. 3, and (e) $B_t$ in Fig. 8. Lower-edge (points $O_1$, $O_3$, $O_5$, $I_1$ and $I_3$) and upper-edge modes (points $O_2$, $O_4$, $O_6$, $I_2$ and $I_4$) of the bandgaps are shown as well.

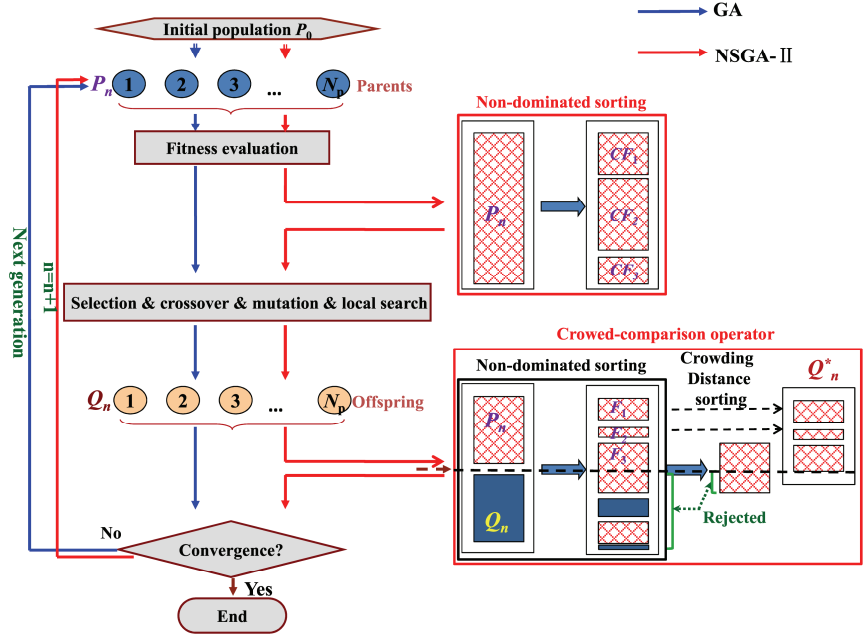

**FIG. 1.** Iterative procedures of GA and NSGA-II.

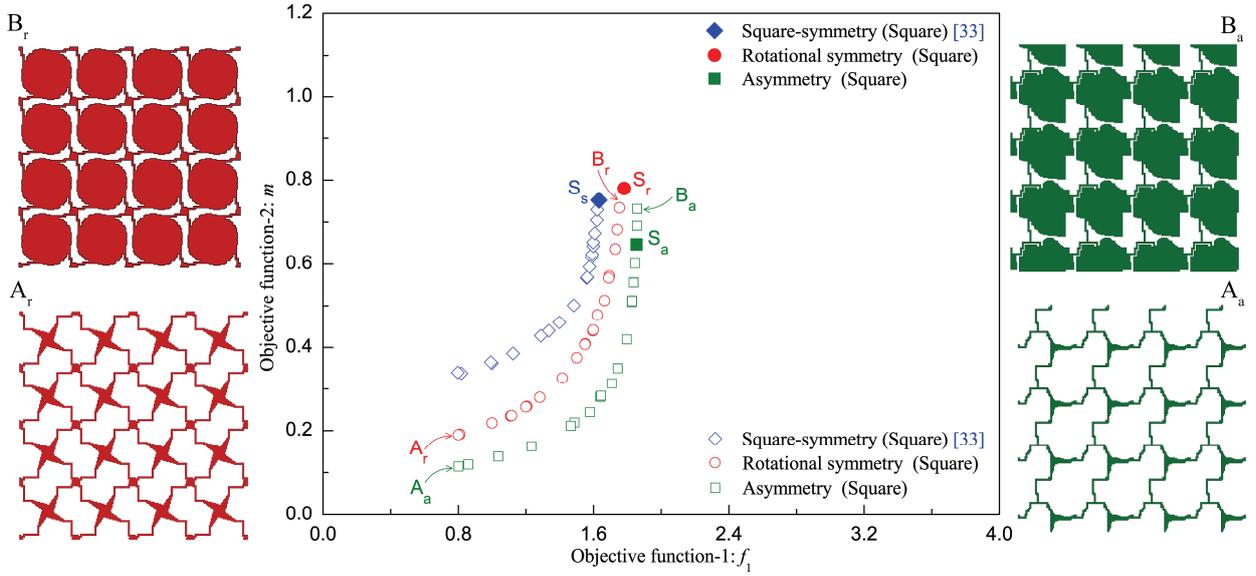

**FIG. 2.** Multi-objective solutions for the square-symmetry (scattered hollow diamonds), rotational symmetry (scattered hollow circles) and asymmetry (scattered hollow squares) in the square lattices with simultaneously maximal relative BGWs and minimal mass for the first bandgap $f_1$ of the out-of-plane wave mode. The results for the square-symmetry are from Ref. [33]. The near-optimal solutions of the SOOP with the square-symmetry (solid diamond), rotational symmetry (solid circle) and asymmetry (solid square) are also shown. The representative designs $A_r$, $B_r$, $A_a$ and $B_a$ are presented as well.

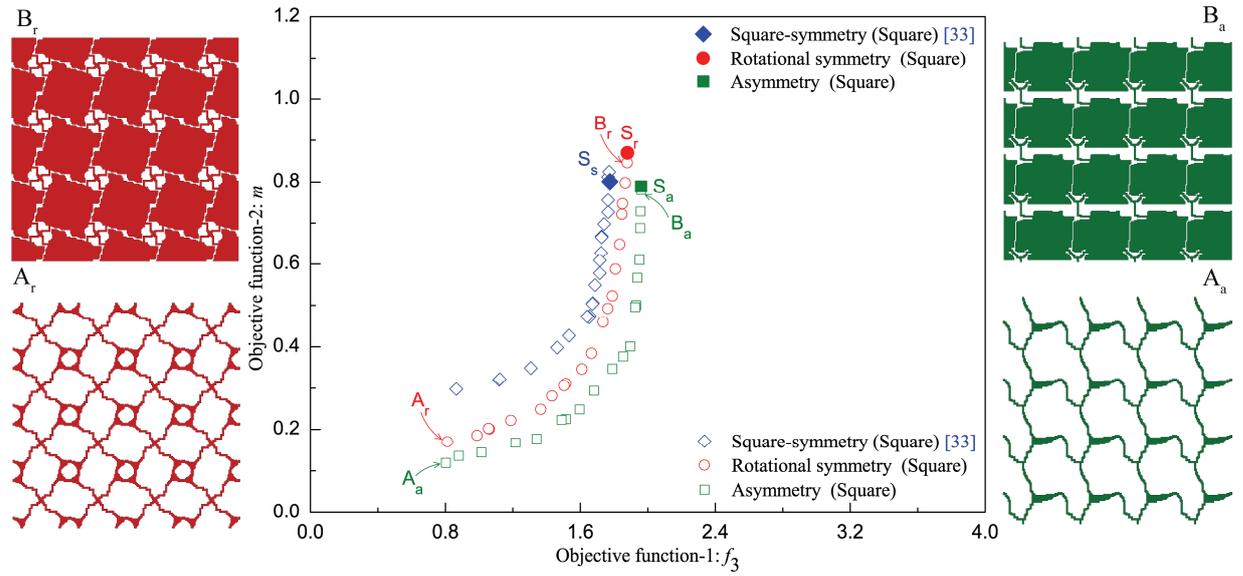

**FIG. 3.** Multi-objective solutions for the square-symmetry (scattered hollow diamonds), rotational symmetry (scattered hollow circles) and asymmetry (scattered hollow squares) in the square lattices with simultaneously maximal relative BGWs and minimal mass for the third bandgap $f_3$ of the in-plane wave mode. The results for the square-symmetry are from Ref. [33]. The near-optimal solutions of the SOOP with the square-symmetry (solid diamond), rotational symmetry (solid circle) and asymmetry (solid square) are also shown. The representative designs $A_r$, $B_r$, $A_a$ and $B_a$ are presented as well.

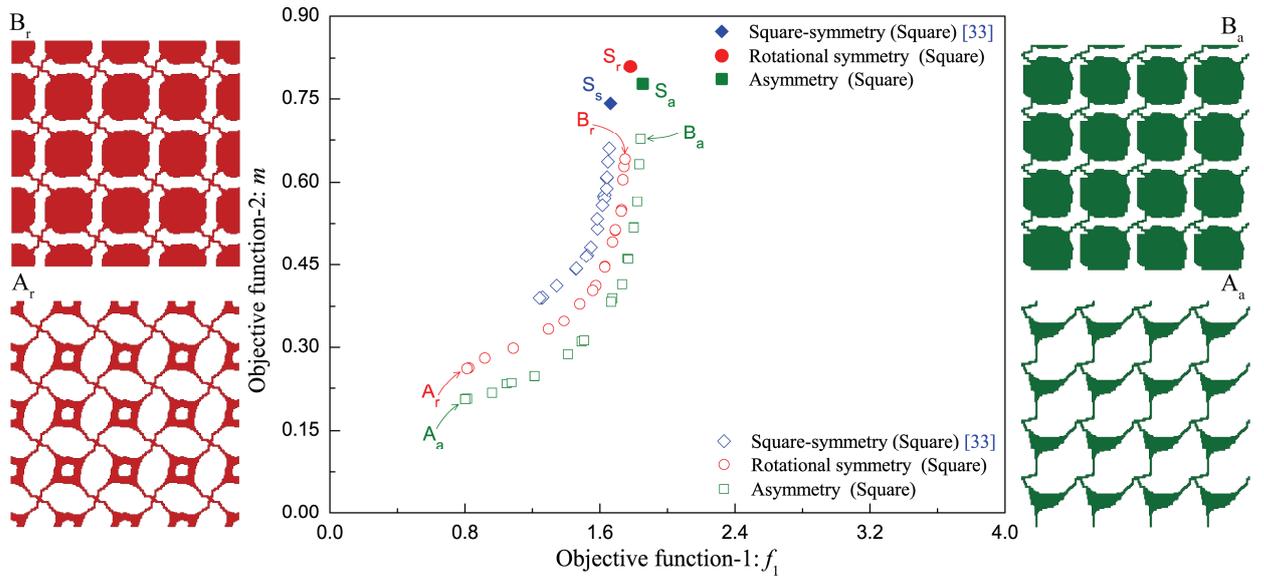

**FIG. 4.** Multi-objective solutions for the square-symmetry (scattered hollow diamonds), rotational symmetry (hollow circles), asymmetry (hollow squares) in square lattices with simultaneously maximal relative BGWs and minimal mass for the first complete bandgap $f_1$ of the full wave modes. The results for the square-symmetry are from Ref. [33]. The near-optimal solutions of the SOOP with the square-symmetry (solid diamond), rotational symmetry (solid circle) and asymmetry (solid square) are also shown. The representative designs $A_r$, $B_r$, $A_a$ and $B_a$ are presented as well.

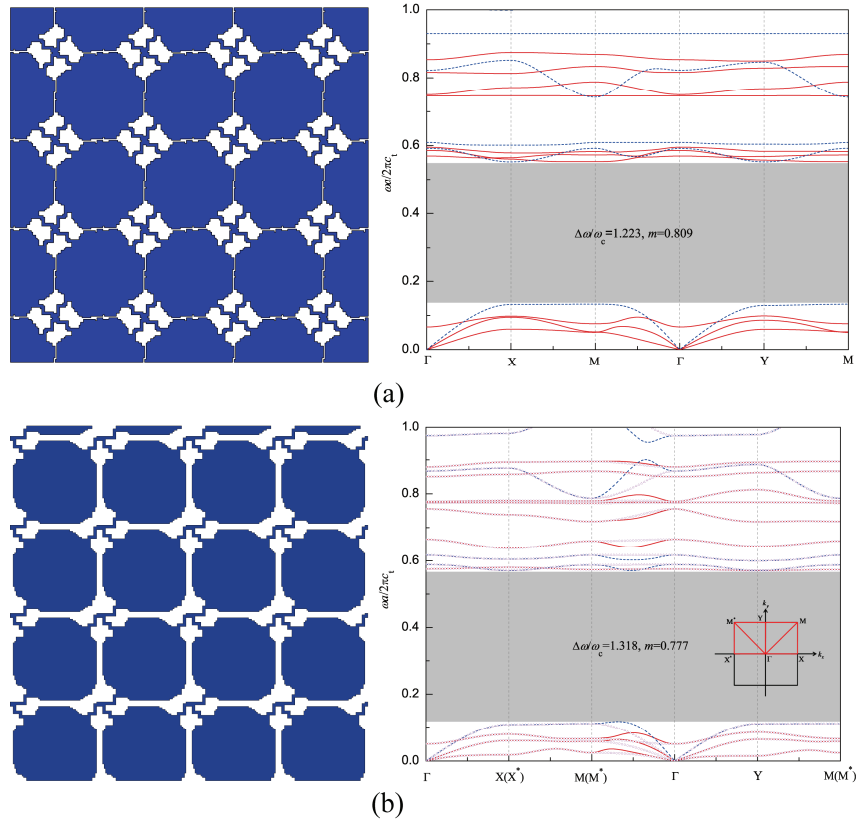

**FIG. 5.** Optimized 4×4 crystal structures $S_r$ (a), $S_a$ (b) in Fig. 4, their band structures and the corresponding first irreducible Brillouin zone. The solid and dashed lines represent the bands for the in-plane and out-of-plane modes, respectively. The normalized frequency $\Omega=\omega a/2\pi c_t$ (with the transverse wave velocity of silicon and the lattice constant $a$) is used.

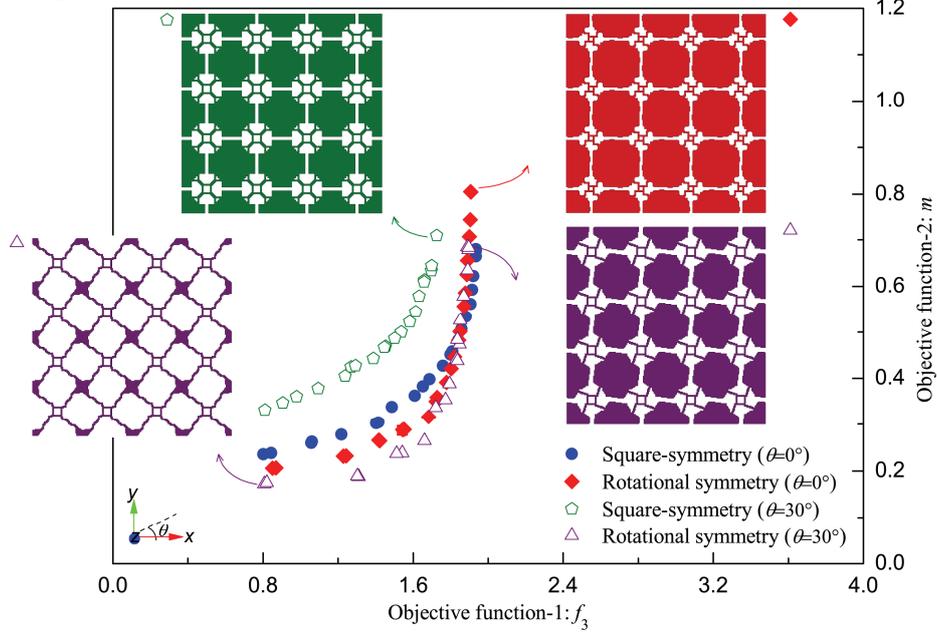

**FIG. 6.** Multi-objective solutions for the square-symmetry (circles and pentagons) and rotational symmetry (squares and triangles) in the square lattices with anisotropic material parameters for the third bandgap $f_3$ of the in-plane wave mode. Different material orientations ($\theta=0°$ and $\theta=30°$) are considered.

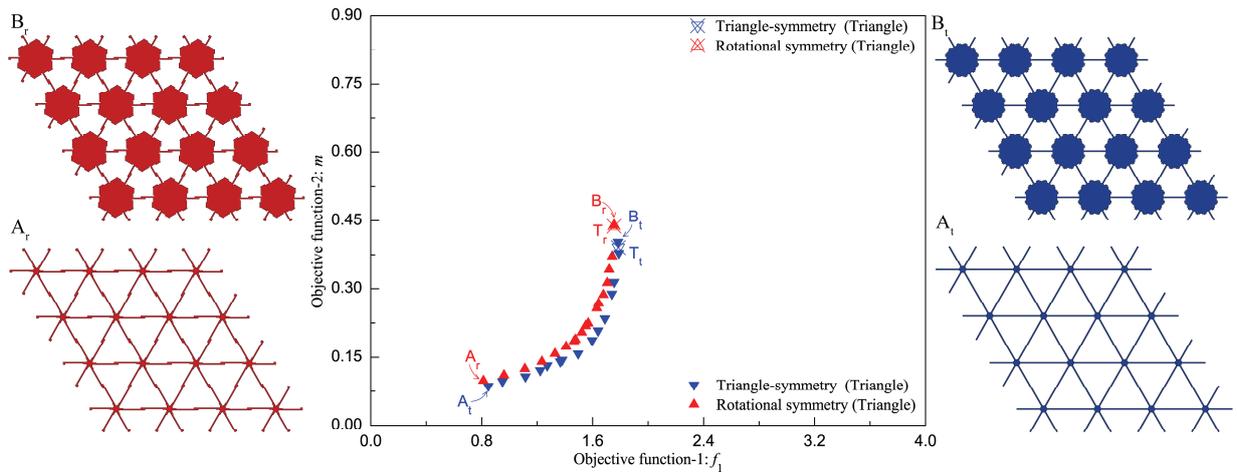

**FIG. 7.** Multi-objective solutions for the triangle-symmetry (solid lower triangles) and rotational symmetry (solid upper triangles) in the triangular lattices with simultaneously maximal relative BGWs and minimal mass for the first bandgap $f_1$ of the out-of-plane wave mode. The near-optimal solutions of the SOOP with the triangle-symmetry (hollow lower triangle with cross) and rotational symmetry (hollow upper triangle with cross) are also shown. The representative designs $A_t$, $B_t$, $A_r$ and $B_r$ are presented as well.

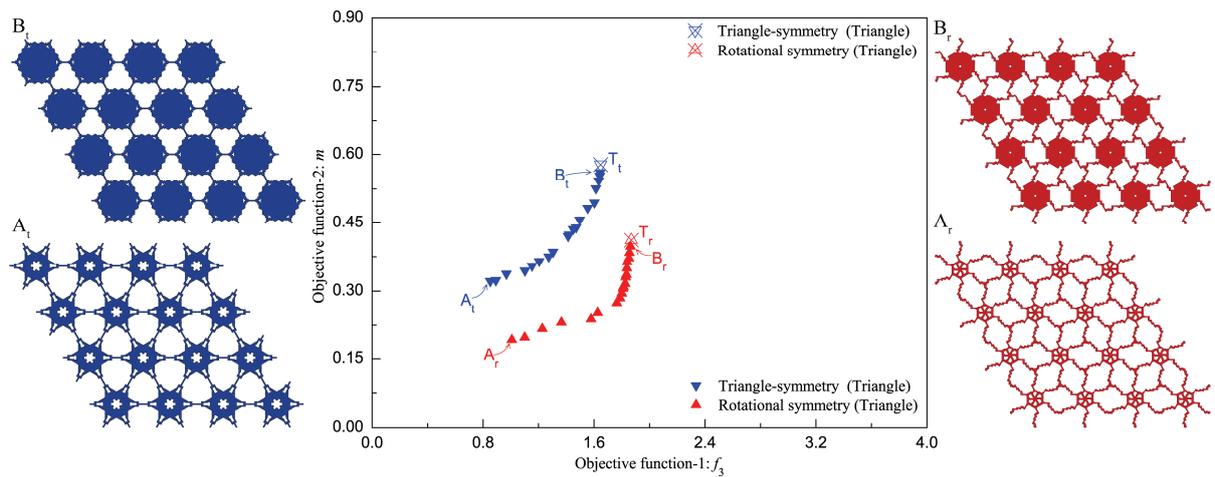

**FIG. 8.** Multi-objective solutions for the triangle-symmetry (solid lower triangles) and rotational symmetry (solid upper triangles) in the triangular lattices with simultaneously maximal relative BGWs and minimal mass for the third bandgap $f_3$ of the in-plane wave mode. The near-optimal solutions for the SOOP with the triangle-symmetry (hollow lower triangle with cross) and rotational symmetry (hollow upper triangle with cross) are also shown. The representative designs $A_t$, $B_t$, $A_r$ and $B_r$ are presented as well.

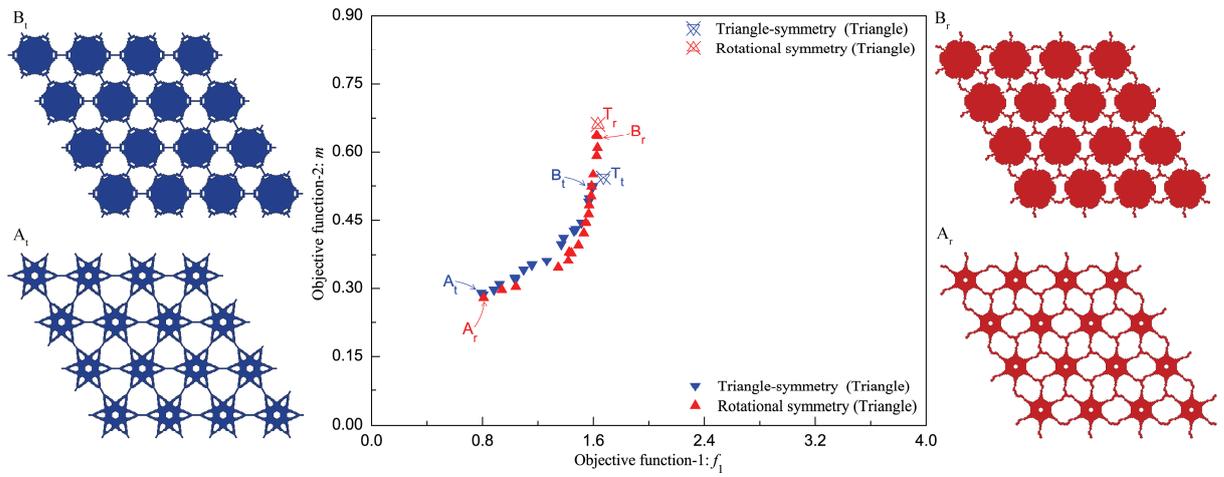

**FIG. 9.** Multi-objective solutions for the triangle-symmetry (solid lower triangles) and rotational symmetry (solid upper triangles) in the triangular lattices with simultaneously maximal relative BGWs and minimal mass for the first complete bandgap $f_1$ of the full wave modes. The near-optimal solutions of the SOOP with the triangle-symmetry (hollow lower triangle with cross) and rotational symmetry (hollow upper triangle with cross) are also shown. The representative designs $A_t$, $B_t$, $A_r$ and $B_r$ are presented as well.

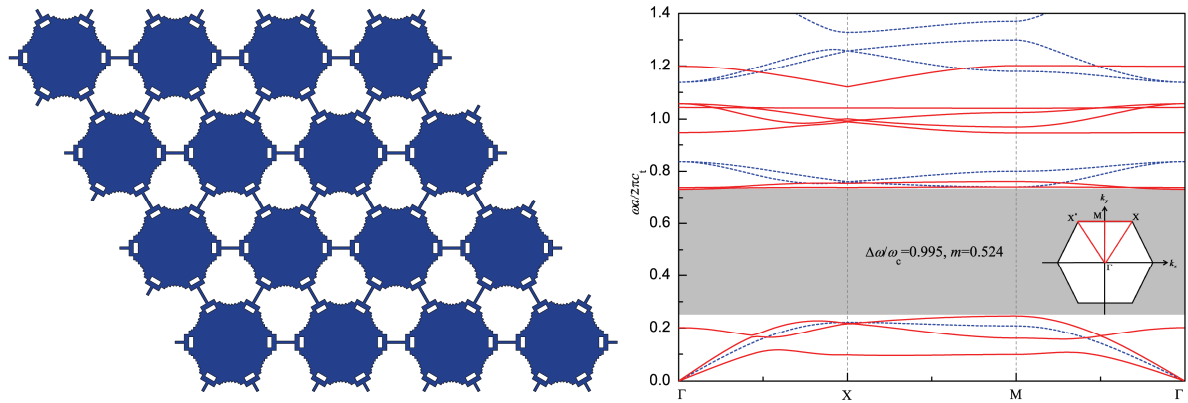

**FIG. 10.** Optimized 4×4 crystal structure $B_t$ in Fig. 9, its band structures and the corresponding first irreducible Brillouin zone.

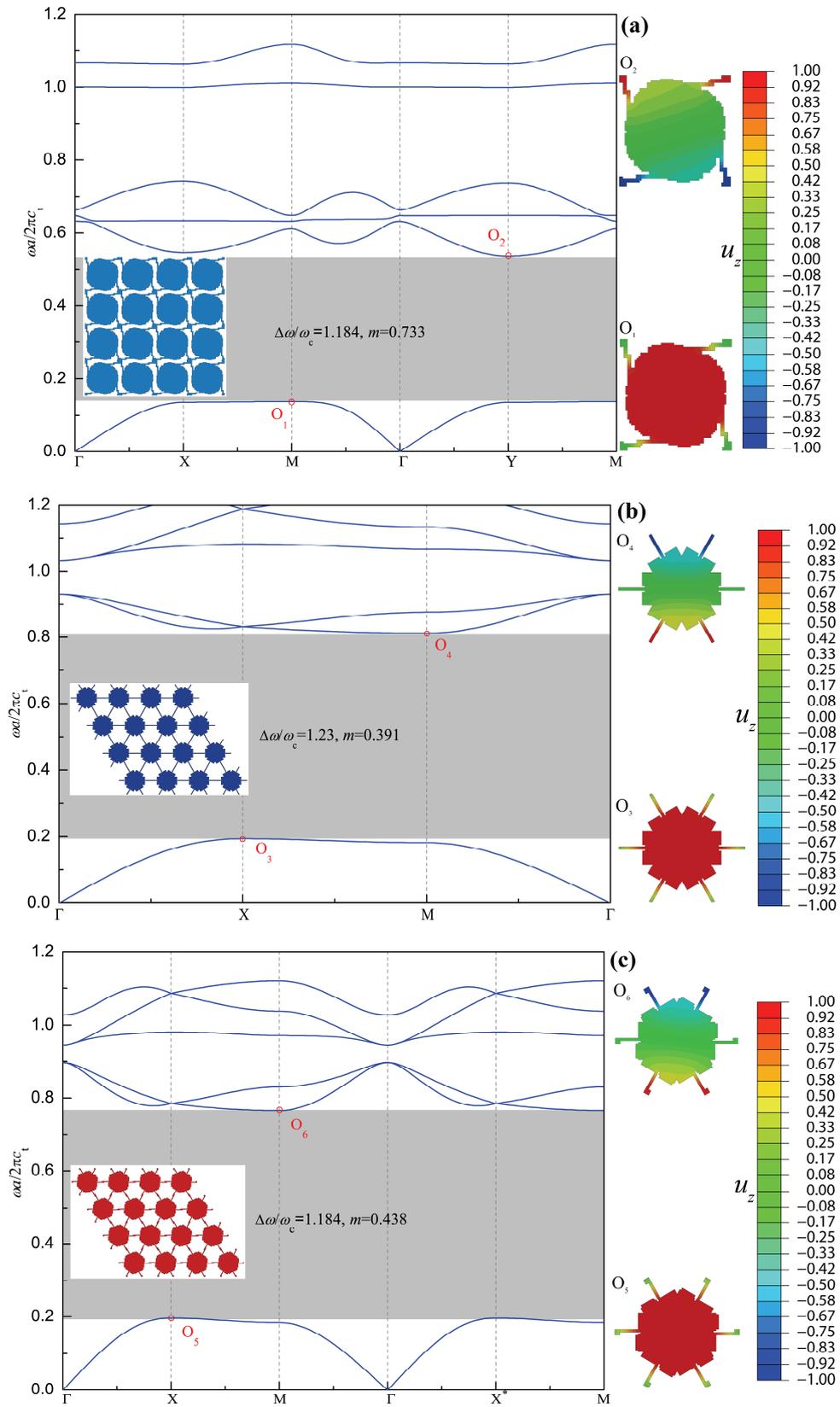

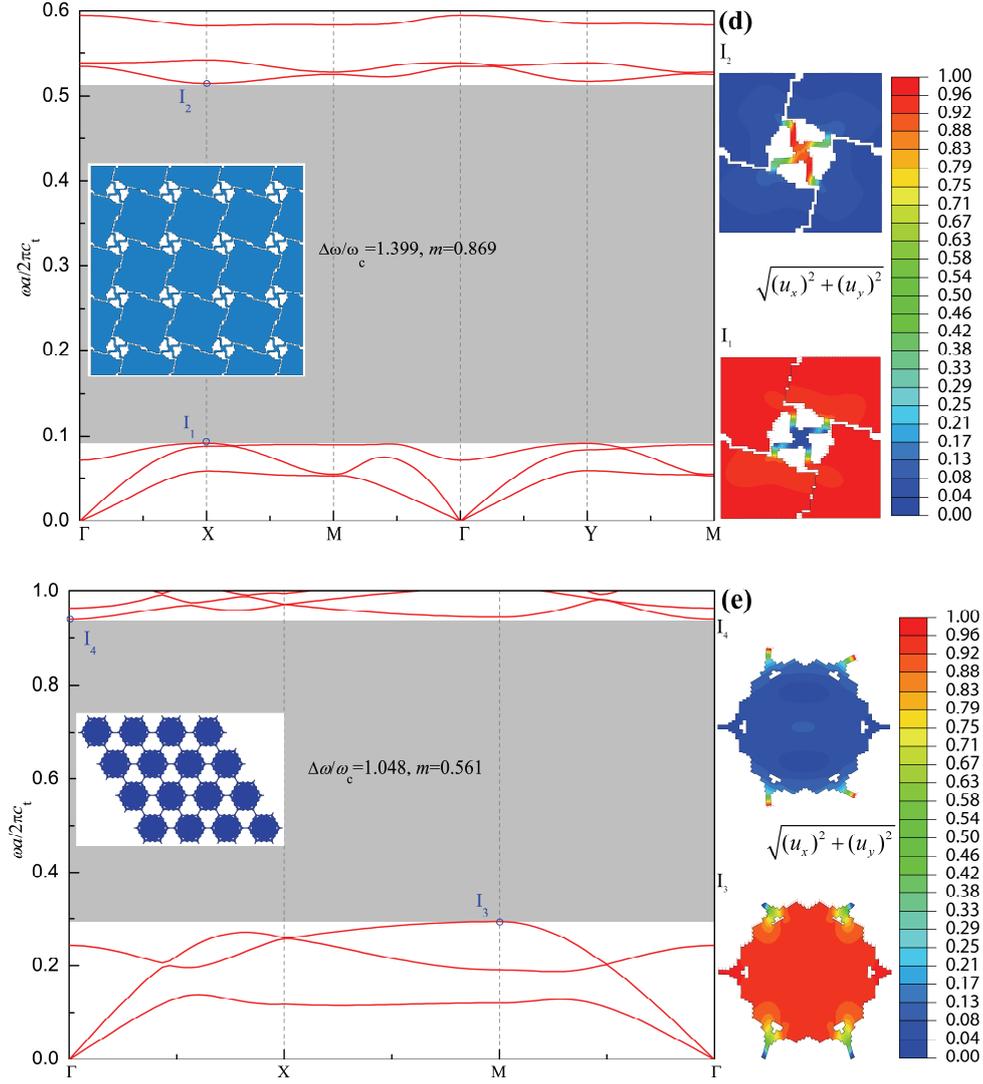

**FIG. 11.** Optimized representative structures and their band structures for the out-of-plane wave: (a) $B_r$ in Fig. 2, (b) $T_t$ in Fig. 7, and (c) $T_r$ in Fig. 7; and in-plane waves: (d) $S_r$ in Fig. 3, and (e) $B_t$ in Fig. 8. Lower-edge (points $O_1$, $O_3$, $O_5$, $I_1$ and $I_3$) and upper-edge modes (points $O_2$, $O_4$, $O_6$, $I_2$ and $I_4$) of the bandgaps are shown as well.